\documentclass[12pt]{article}
\usepackage[tbtags]{amsmath}
\usepackage{amssymb,mathabx,textcomp,color,times,setspace, cleveref,graphicx,float}
\pdfoutput=1
\usepackage[hidelinks,bookmarks=false]{hyperref}
\usepackage[page,title,titletoc,header]{appendix}
\usepackage[top=1in, bottom=1in, left=1in, right=1in]{geometry}
\usepackage{url,indentfirst,comment,tikz,titlesec,secdot,enumerate,psfrag,epsf}
\usepackage[style=authoryear-comp,sortcites=false,firstinits=true, backend=bibtex,natbib=true,hyperref=true,maxbibnames=99,maxcitenames=2]{biblatex}
\renewbibmacro{in:}{}
\usetikzlibrary{arrows}
\usetikzlibrary{shapes}
\usepackage{xy}\xyoption{all} \xyoption{poly} \xyoption{knot}
  \author{Olanrewaju Akande, Fan Li and Jerome Reiter \footnote{Olanrewaju Akande is PhD student, Department of Statistical Science, Duke University, Durham, NC 27708 (E-mail: \href{mailto:olanrewaju.akande@duke.edu}{olanrewaju.akande@duke.edu}); Fan Li is Associate Professor, Department of Statistical Science, Duke University, NC 27708 (E-mail: \href{mailto:fli@stat.duke.edu}{fli@stat.duke.edu}); and Jerome P. Reiter is Professor of Statistical Science, Duke University, Durham, NC 27708 (E-mail: \href{mailto:jerry@stat.duke.edu}{jerry@stat.duke.edu}). This research was supported by grants from the National Science Foundation (SES-11-31897) and the Alfred P. Sloan Foundation (G-2015-20166003).}  }
  \title{An Empirical Comparison of Multiple Imputation Methods for Categorical Data}
  \date{}
\newcommand*\circled[1]{\tikz[baseline=(char.base)]{
            \node[shape=circle,draw,inner sep=5pt] (char) {#1};}}

\titleformat*{\section}{\large\bfseries\centering}
\titleformat*{\subsection}{\normalsize\bfseries}	

\usepackage[labelsep=period]{caption}
\usepackage{subcaption}
\begin{document}
\maketitle

\begin{abstract}
\noindent Multiple imputation is a  common approach for dealing with
missing values in statistical databases. The imputer fills in
missing values with draws from predictive models estimated from the
observed data, resulting in multiple, completed versions of the
database.  Researchers have developed a variety of default routines to
implement multiple imputation; however, there has been
limited research comparing the performance of these methods,
particularly for categorical data.  We use simulation studies to
compare repeated sampling properties of three default multiple imputation
methods for categorical data, including chained equations using generalized linear models, chained equations
using classification and regression trees, and a fully Bayesian joint
distribution based on Dirichlet Process mixture models. We base the
simulations on categorical data from the American Community Survey.
In the circumstances of this study, the results suggest that default chained equations approaches based on
generalized linear models are
dominated by the default regression tree and Bayesian mixture model approaches.  They
also suggest competing advantages for the regression tree and Bayesian mixture
model approaches, making both reasonable default engines for multiple
imputation of categorical data. A supplementary material for this article is available online.
\end{abstract}


\section{INTRODUCTION}
Nearly all sample surveys and censuses suffer from item nonresponse, e.g.,
individuals do not answer some questions.  It is well-known that statistical analyses based
on only the complete cases (with all variables observed) or available
cases (with all variables observed for the specific
analysis) can be problematic.  At best, such analyses are
inefficient, as they sacrifice information from partially observed
responses. At worst, they result in biased inferences when there are
systematic differences between the observed data and the missing data
\citep{Rubin1976}. See \citet{LittleRubin2002} for additional
discussion of the drawbacks of complete/available case analysis.

A common approach to handling item nonresponse is multiple imputation
(MI) \citep{Rubin1987}. In MI, the analyst creates multiple, completed
datasets by replacing the missing values with draws from
(posterior) predictive distributions estimated with the observed data.  The analyst estimates
quantities of interest using each completed data set, and combines the
estimates using methods developed by \citet{Rubin1987}. This process
incorporates the additional uncertainty from the missing data in
inferences. The completed datasets also can be released as
public use files. For reviews of MI, see
\citet{Rubin1996,Schafer1997a,BarnardMeng1999, ReiterRaghunathan2007, HarelZhou2007}.

When implementing MI, most analysts adopt one of two
general classes of strategies: joint modeling (JM) and fully
conditional specification (FCS)
\citep{Buuren2007}. In the JM strategy, the analyst specifies a joint
distribution for all variables in the data. Imputations are sampled
from the implied conditional distributions of the variables with
missing data, given all other variables. The JM strategy is appealing,
in that it accords with the theory in \citet{Rubin1987}. In
practice, however, specifying accurate joint distributions for large numbers of
variables can be challenging. Indeed, most popular JM approaches --- such as AMELIA
\citep{HonakerEtAl2011}, proc MI in SAS \citep{Yuan2011}, and the routine ``norm'' in R \citep{Schafer1997a} --- make a simplifying assumption that the data follow
multivariate Gaussian distributions, even for categorical variables. 
Another option for categorical data is to use a log-linear model; this is
implemented in the R package ``cat'' \citep{Schafer1997a}.

In the FCS strategy, the analyst directly specifies and samples from
the univariate  distributions of each variable conditional on all
other variables, without first forming a proper joint distribution.
The most popular FCS approach is known as multiple imputation by chained
equations (MICE), which uses generalized linear models for each
conditional distribution \citep{RaghunathanEtAl2001,
  BuurenGroothuis-Oudshoorn2011, RoystonWhite2011, SuEtAl2011}.
The main strength of MICE lies in its simplicity and
flexibility --- one can tailor the predictive model for individual
variables, e.g., use a logistic regression for binary variables and a
linear regression for continuous variables.  However, MICE lacks
theoretical basis in that the specified univariate conditional
distributions may not be compatible \citep{ArnoldPress1989,
  GelmanSpeed1993}; that is, the set of the conditional distributions
may not correspond to any joint distribution. Therefore, MICE defines
a potentially incompatible Gibbs sampler
\citep{LiEtAl2014}.  Despite this theoretical drawback,
simulation-based research suggests that MICE performs
well in practice \citep[][to name a
few]{Brand1999,BuurenEtAl1999,RaghunathanEtAl2001,Rubin2003,
  BuurenEtAl2006, Buuren2007}. Software packages implementing MICE
include ``IVEware'' in SAS \citep{RaghunathanEtAl2002}, ``mice'' \citep{BuurenGroothuis-Oudshoorn2011, Buuren2012} and ``mi''
in R \citep{SuEtAl2011}, and ``mi'' and ``ice" in STATA \citep{RoystonWhite2011}.

When the data include a large number of exclusively categorical
variables, analysts implementing MI face a challenging task.
When using log-linear models for a JM approach, the space of possible models is enormous due to the large
number of potential interaction effects, making it difficult to
select an imputation model.  Similar model selection issues plague MICE
approaches, e.g., when specifying a multinomial logistic regression
model for a multi-valued variable with missing data. Many analysts
default to including main effects only; indeed, this is the default
option in most popular MICE packages.

Motivated by these shortcomings, several authors have developed more
flexible, default MI engines for both strategies.  For JM imputation,
\citet{SiReiter2013} and \citet{Manrique-VallierReiter2014b,Manrique-VallierReiter2014a} use
Dirichlet process mixture of products of multinomial distributions
(DPMPM), which are nonparametric Bayesian versions of the latent class
models used by \citet{VermuntEtAl2008}. The DPMPM imputation routines
are implemented in the R
software package, ``NPBayesImpute'' \citep{Manrique-VallierEtAl2014}.  For FCS imputation,
\citet{BurgetteReiter2010} and \citet{Buuren2012} use classification
and regression trees (CART). MI via trees is available as an option in
``mice'' in R.  Independently, these techniques have shown
promise as general-purpose routines for MI, suggesting that
they may be preferable to default applications of MICE. However, to our knowledge,
these two techniques have never been compared in
simulation contexts with large-dimensional, genuine categorical data,
nor have they been compared to default implementations of chained equations with
generalized linear models.   These lack of comparisons make it
difficult, if not impossible, to assess the relative merits of each
procedure.

In this article, we compare these three
MI methods for categorical data using repeated sampling
simulations. Using hypothetical populations comprising data from the
American Community Survey (ACS), we find that the DPMPM and CART imputation
engines (which we label with MI-DPM and MI-CART, respectively) result
in better repeated sampling performance than the standard chained
equations via generalized linear models (which we label with
MI-GLM). The results suggest competing advantages for the MI-CART and
MI-DPM approaches, depending on the sample size and amount of missing
data.  Both procedures are sensible default choices for
MI of categorical data.

The remainder of this article is organized as follows. In Section 2,
we review  the three methods for MI with categorical
data. In  Section 3, we describe the simulation design and results
with ACS data.  In Section 4, we conclude with general lessons learned and
some suggestions for practical implementation.

\section{MULTIPLE IMPUTATION FOR CATEGORICAL DATA}

We first introduce notation for describing MI in the context of
categorical data. Let $Y_{ij} \in \{1, \dots, D_j\}$ be the value of
variable $j$ for individual $i$, where $j=1, \dots, p$ and $i=1,
\dots, n$. For each individual $i$, let ${\bf Y}_i =
(Y_{i1}, \dots, Y_{ip})$.  Let ${\bf Y} = ({\bf Y}_1, \ldots, {\bf
  Y}_n)$ be the $n \times p$ matrix comprising the data for all
records included in the sample.  We write ${\bf Y} = ({\bf
  Y}_{\textrm{obs}}, {\bf Y}_{\textrm{mis}})$, where ${\bf
  Y}_{\textrm{obs}}$ and ${\bf Y}_{\textrm{mis}}$ are respectively the
observed and missing parts of ${\bf Y}$.  We write ${\bf
  Y}_{\textrm{mis}} = ({\bf Y}_{\textrm{mis}1}, \dots, {\bf
  Y}_{\textrm{mis}p})$, where ${\bf Y}_{\textrm{mis}j}$ represents all
missing values for variable $j$, where $j=1, \dots, p$.  Similarly, we write ${\bf
  Y}_{\textrm{obs}} = ({\bf Y}_{\textrm{obs}1}, \dots, {\bf
  Y}_{\textrm{obs}p})$ for the corresponding observed data.

In MI, the analyst generates values of ${\bf Y}_{\textrm{mis}}$ using
models estimated with ${\bf Y}_{\textrm{obs}}$. This results in a
completed dataset  ${\bf Y}^{(l)}$. The analyst draws $L$ completed
datasets, $\{{\bf Y}^{(l)}: l
= 1, \dots, L\}$, that are available for dissemination or analysis.

\subsection{Overview of MI-GLM}

In MI-GLM, we generate imputations from a sequence of  predictive
distributions derived from univariate generalized linear models.  The
approach proceeds as follows.  We first specify an order for imputing
the variables.  For example, the ``mice'' package in R by default imputes the
variables in the order that they appear in the data matrix (although a different order can be specified by the user), and the ``IVEWare'' package imputes
variables in increasing order of the number of missing cases. Suppose that $r_0 \leq p$ variables have missing values, and
that $r_1 = p - r_0$ variables are fully complete.  
Let the ordered list of variables be defined as $(Y_{(1)},
\dots, Y_{(p)})$, where any variable with no missing values is put at the end
of the list. We next fill in initial imputations at the missing values.  These
can be obtained from random draws from
the observed marginal distributions for each variable with missing data (the ``mice'' package in R does this).
Alternatively, for each variable $j$ with missing values, we can sample from conditional distributions, $(Y_{(j)}
\mid Y_{(1)}, \dots, Y_{(j-1)}, Y_{(r_0+1)}, \dots, Y_{(p)})$, where
each model is estimated on the available cases for that
model.

With this initial set of completed data, we now use an iterative process akin to a Gibbs sampler to update the imputations. At each iteration $t$ of the updating, we estimate the predictive model, $(Y_{(1)} \mid {\bf Y}_{\textrm{obs}(1)}, \{\bf Y\}^{(t-1)}_{(k)}: k > 1\})$, where ${\bf Y}_{(k)}^{(t-1)}$ includes the set of observed and imputed
values for variable $k$ at iteration $t-1$.  We replace ${\bf
  Y}_{\textrm{mis}(1)}^{(t-1)}$ with draws from this conditional
distribution, ${\bf Y}_{\textrm{mis}(1)}^{(t)}$. We repeat this process for each variable $j$ with
missing data, estimating and imputing from each predictive model, $(Y_{(j)} \mid {\bf
  Y}_{\textrm{obs}(j)}, \{{\bf Y}^{(t)}_{(k)}: k < j\}, \{{\bf
  Y}^{(t-1)}_{(k)}: k > j\})$. 
 We repeat the cycle $t>1$ times. 
The values at the final iteration are used to create the completed
dataset, $\mathbf{Y}^{(l)} = ({\bf
  Y}_{\textrm{obs}}, {\bf Y}_{\textrm{mis}}^{(l)})$.  The entire
process is replicated $L$ times to create the full set of multiple
imputations.

For categorical data, most implementations of MI-GLM use logistic
regressions for binary variables, multinomial logistic regressions for
unordered variables, and cumulative logistic regressions
for ordered variables.  The default specification of the predictor
functions is to include main effects only for all $p-1$ variables.  It
is possible to remove predictors from and add interactions to certain univariate conditionals using special commands; for ``mice'' in R, the predictorMatrix argument can be used to add or remove predictors and the passive imputation option can be used to add interactions. For $Y_{(j)}$ corresponding 
to unordered categorical variables with large numbers of levels, multinomial logistic regressions can have large
numbers of parameters, which can result in unstable coefficient estimates with high standard errors.

\subsection{Overview of MI-CART}
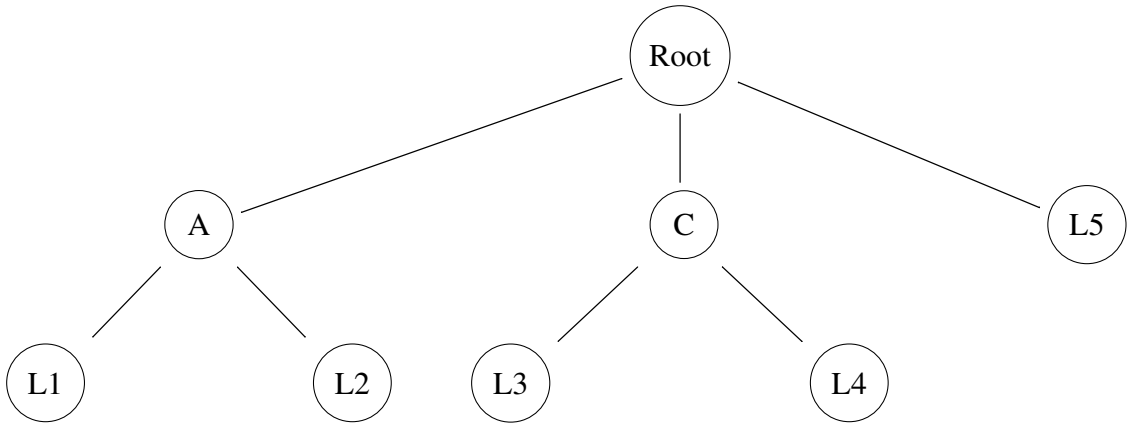
\begin{figure}
    $$ \vcenter{\xymatrix{
    & & & & \textrm{\circled{Root}}  \ar@{-}[dlll] \ar@{-}[d] \ar@{-}[drrr] & & & & \\
    & \textrm{\circled{A}} \ar@{-}[dl] \ar@{-}[dr] & & & \textrm{ \circled{C}} \ar@{-}[dl] \ar@{-}[dr] & & & \textrm{\circled{L5}} & \\
    \textrm{\circled{L1}} & & \textrm{\circled{L2}} & \textrm{\circled{L3}} & & \textrm{\circled{L4}} & & & }} $$
    \caption{Illustration of the tree structure in CART. A:
      African-Americans; C: Caucasian; H: Hispanic; M: male; F:
      female. Leaf L1 contains female
African-Americans; leaf L2 contains male African-Americans; leaf L3
contains female Caucasians; leaf L4 contains male Caucasians; and,
leaf L5 contains Hispanics of both genders.}
    \label{fig:CART}
    \end{figure}

CART for categorical data \citep{BreimanEtAl1984}, originally
developed for classification, can be modified to be an imputation
engine.  For any outcome $Y_{(j)}$, CART recursively partitions the predictor space in a way that can
be effectively represented by a tree structure, with leaves
corresponding to the subsets of units. The values of $Y_{(j)}$ in each leaf
can be conceived as draws from the conditional distribution of
$Y_{(j)}$ for the set of predictor values that satisfy the partitioning criteria that
define that leaf. To illustrate, Figure \ref{fig:CART} displays a
fictional regression tree
for an outcome variable and two predictors, gender (male or female)
and race/ethnicity (African-American, Caucasian, or Hispanic).  To
approximate the conditional distribution of $Y_{(j)}$ for a particular
gender and race/ethnicity combination, one uses the values in the corresponding
leaf. For example, to approximate
the distribution of $Y_{(j)}$ for female Caucasians, one uses the
observed values of the outcome in leaf L3.

MI-CART operates like MI-GLM, except that CART models are used in
place of logistic regressions.  Initial imputations can be obtained by
running CART on the available cases for $(Y_{(j)} \mid Y_{(1)}, \dots,
Y_{(j-1)}, Y_{(r_0+1)}, \dots, Y_{(p)})$.  For any value of
$(Y_{i(1)}, \dots, Y_{i(j-1)}, Y_{i(r_0+1)}, \dots, Y_{i(p)})$, including values for
previously initialized variables, we sample the initial imputation by dropping down the tree for
$Y_{(j)}$ until finding the appropriate leaf, and sample from the
values in that leaf. \citet{BurgetteReiter2010} suggest an additional step before 
sampling values within the leaves, namely to use Bayesian bootstraps
within each leaf (holding constant the number of observations in each
leaf) 
to create the pools from which to sample.  The Bayesian bootstrap incorporates uncertainty about the distributions in each leaf.  As with MI-GLM, the process is cycled
$t>1$ times, resulting in a completed dataset.   We note, however, that this imputation routine 
does not incorporate uncertainty about the split points in the trees.

CART algorithms, and hence MI-CART, can struggle with
categorical predictors with many levels. For example, a categorical
predictor with 32 levels results in over two billion potential
partitions, making it computationally difficult to run the algorithm
in sensible time with standard computers. 

\subsection{Overview of MI-DPM}
The MI-DPM procedure assumes that the distribution of the categorical
data can be characterized using a latent class model. We assume that all combinations of variables are possible {\it a priori} -- that is, there are no structural zeros \citep{SiReiter2013}.   Each individual
belongs to one of $K < \infty$ latent classes.  Within
each class, all variables follow independent multinomial
distributions.  To express this as a formal probability model, let
$z_i \in (1, \dots, K)$ represent the latent class of individual $i$; $\pi_k = \Pr(z_i = k)$, for $k = 1, \dots, K$, be the probability of being in latent class $k$; $\lambda_{kjy} =
\Pr(Y_{ij} = y \mid z_i = k)$ be the probability that variable $Y_{ij}$
takes on the value $y$ for records in latent class $k$.  Also, let
$\boldsymbol{\pi} = (\pi_1,\dots, \pi_K)$, and let
$\boldsymbol{\lambda} =  \{\lambda_{kjy} : k =1, \dots, K; j = 1,
\dots, p; y = 1, \dots, D_j\}$. We have
    \begin{eqnarray}
    Y_{ij} \mid z_i, \boldsymbol{\lambda} &\overset{indep}{\sim}&
    \textrm{Discrete}_{1:D_j} (\lambda_{z_ij1}, \dots,
    \lambda_{z_ijD_j})  \ \ \ \textrm{for all $i$ and $j$} \label{eq1}\\
    z_i \mid \boldsymbol{\pi} &\overset{iid}{\sim}&
    \textrm{Discrete}_{1:K} (\pi_1, \dots, \pi_K)  \ \ \ \textrm{for
      all $i$} \label{eq2}.
    \end{eqnarray}
We can express the marginal probability of any quantity by averaging
over the latent classes; for example,
\begin{equation}
   Pr(Y_{i1} = y_1,\ldots,Y_{ip} = y_p \mid \boldsymbol{\lambda}, \boldsymbol{\pi}) = \sum\limits_{k=1}^{K} \pi_k \prod\limits_{j=1}^{p} \lambda_{kjy_j}.
    \end{equation}
Although variables are independent within classes, averaging over
classes can result in dependence among the variables.
\citet{DunsonXing2009} show that, with large enough $K$, this model is
consistent for any joint probability distribution.

For prior distributions, \citet{SiReiter2013} and
\citet{Manrique-VallierReiter2014a} use a truncated version of the stick-breaking
representation of the Dirichlet process proposed by
\citet{DunsonXing2009},
    \begin{eqnarray}
    (\lambda_{kj1}, \dots, \lambda_{kjD_j}) &\overset{indep}{\sim}& \textrm{Dirichlet}({\bf 1}_{D_j})\\
    \pi_k &=& V_k \prod\limits_{h<k}^{} (1 - V_h) \\
     V_k &\overset{iid}{\sim}& \textrm{Beta} (1, \alpha) \ \  \textrm{for} \ \ k = 1, \ldots, K - 1; V_K = 1\\
    \alpha &\sim& \textrm{Gamma} (0.25, 0.25).
    \end{eqnarray}
This prior distribution facilitates efficient computation while
allowing the data to dominate the posterior distribution. 

To set $K$, we recommend the approach in \citet{SiReiter2013} and
\citet{Manrique-VallierReiter2014a}.  The analyst  starts with a modest value, say $K = 20$.  In each iteration of the MCMC, the 
analyst can compute the number of classes that have at least one individual.  If the number of occupied classes reaches $K$ across
any of the iterations, then the analyst should increase $K$ and repeat.  As long as $K$ is large enough, the estimated 
posterior distribution is largely insensitive to the value of $K$.  The iterative method of selecting $K$ primarily serves 
to improve computational efficiency, as it avoids estimating parameters for many empty (and essentially irrelevant) classes.

Posterior inferences are obtained via a Gibbs sampler.  Missing
values are handled within the sampler. Given a draw of the parameters
and observed data, one samples a value of the latent class indicator for
the record using \eqref{eq2}.  Given a draw of the latent class indicator, one samples
values for the missing items using independent draws from \eqref{eq1}.  The independence of variables
within the latent classes facilitates computation and imputation in
datasets with many categorical variables. The MI-DPM  approach can be
adapted for categorical data with structural zeros
\citep{Manrique-VallierReiter2014b}, and it is not limited to a maximum number of levels.  Because it relies on a coherent joint model, the MI-DPM engine can be the building block for handling more
complicated missing data situations, including nonignorable nonresponse \citep{SiEtAl2015} and  measurement error
\citep{Manrique-VallierReiter2015}.

\section{REPEATED SAMPLING EVALUATIONS} \label{rep_samp_eva}

We base the simulation studies on data from the public use microdata files from
the 2012 ACS, available for download from the United States Bureau of the Census
(\url{http://www2.census.gov/acs2012_1yr/pums/}). As brief background, the ACS 
is sent to about 1 in 38 households in the United States, with the
goal of enabling estimation of population demographics and housing characteristics
at small geographies, e.g., at census tracts.  
For each sampled household, the survey includes questions about the individuals living in the household
 (e.g., their ages, races, incomes) and about the characteristics of
 the housing unit (e.g., number of bedrooms,
 presence of running water or not, presence of a telephone line or not).  The public
 use data include about 1.5 million housing units. 

To construct the population data for the simulation, we treat the
housing unit as the unit of analysis. We disregard data on individual
people  but keep
ACS summaries of family compositions that can be viewed as household-level variables (e.g., presence
of children or not, number of workers in household).  We remove all
identification variables (serial number, region, state, area code and division),
variables corresponding to allocation flags (binary variables indicating if the corresponding variables have been imputed),
variables corresponding to replicate survey weights, and all continuous variables. To simplify comparisons, we eliminate structural zeros --- i.e.,
impossible combinations of categories like a married 12-year old --- by
removing variables or levels.  We also delete about 830,000 units
corresponding to vacant houses and households with single occupants,
again to avoid structural zeros. The final data comprise 671,153 housing units and
35 categorical variables. We treat these as a population from which to sample. We treat all variables as unordered categorical variables to simplify imputation modeling and analysis. The data comprise 11 binary variables (seven of which have marginal probabilities very close to one or zero), 17 variables with three to nine levels, and 7 variables with ten or more levels. The variables
are described in Table 1 in the supplementary material.

We simulate the processes of random sampling and nonresponse
repeatedly, so as to examine frequentist properties of the three
imputation procedures.  We take samples of size $n \in
\{1000, 10000\}$ and randomly blank either 30\% or 45\% of the values
of each variable independently across variables.
This results in data that are missing completely at random (MCAR), which is
the most favorable scenario for all MI procedures.  We also create a
missing at random (MAR)  scenario with $n=10000$ and a 30\%
missingness rate. In each
sample, we use MI-GLM, MI-CART, and MI-DPM to create three sets of
$L=10$ completed datasets. We repeat the process 200 times, each time
generating new samples and new missingness patterns.  We note that the missing data mechanism tends to result in datasets with 10 or fewer 
cases with all variables measured, making complete case analysis untenable.

To implement both MI-GLM and MI-CART, we use the ``mice'' package in R
\citep{BuurenGroothuis-Oudshoorn2014}. For MI-GLM, we use the default choices for binary and nominal
variables with two or more levels: logistic regressions with main effects for all variables. For MI-CART (the ``cart'' option
in ``mice''), we use the default arguments for fitting the trees, e.g.,
at least four observations in any terminal node. We run both MI-GLM
and MI-CART for $t=10$ cycles  as suggested in \citet{BuurenGroothuis-Oudshoorn2011}.  To implement the DPMPM,
we use the  ``NPBayesImpute'' package in R developed by
\citet{Manrique-VallierEtAl2014}. We set the number of latent classes
$K=35$, which appears sufficiently large based on tuning with initial runs. We
run each MCMC chain for $10000$ iterations using the first $2000$ as burn-in.

\subsection{Performance Measures}

We evaluate the MI methods using sets of marginal probabilities,
bivariate probabilities, and trivariate probabilities. We 
consider only estimands that satisfy  $np>10$ and $n(1-p)>10$, where
$p$ is the probability in the population, to eliminate estimands where
the central limit theorem is not likely to hold even with no missing data. For each
estimand, we use each set of $L$ completed datasets to compute point
and interval estimates via the methods of \citet{Rubin1987}. As a brief
review, let $q$ be the completed-data point estimator of some estimand $Q$, and let
$u$ be the estimator of variance associated with $q$.  For $l=1,
\dots, L$, let $q_l$ and $u_l$ be the values of $q$ and $u$ in
completed dataset $\mathbf{Y}^{(l)}$.  We use $\bar{q} =
\sum_{l=1}^L q_l/L$ as the point estimate of $Q$.  We use $T = (1 +
1/L)b + \bar{u}$ as the estimated variance of $\bar{q}$, where $b =
\sum_{l=1}^L (q_l - \bar{q})^2/(L-1)$ and $\bar{u} = \sum_{l=1}^L
u_l/L$.  We form 95\% confidence intervals using $(\bar{q} - Q) \sim
t_{v}(0, T)$, where $t_{v}$ is a $t$-distribution with $v = (L-1)(1 + \bar{u}/ [(1+1/L) b])^2$ degrees of freedom.
Multiple imputation inferences require that 
(i) the central limit theorem applies for the complete-data estimate, and (ii) the sampling distribution of the complete-data point estimates is 
approximately normal. When $p$ is near zero or one, these conditions may not hold, which can result in unreliable inferences including intervals that include zero or one.

For each estimand, we compare the three MI procedures by two metrics.
We also compute these metrics for estimates based on the sampled data before introduction of missing values.  First, we compute
the proportion of the two hundred 95\% confidence intervals that contain the corresponding $Q$ from the full ACS data. Second, we compute the relative mean squared error (Rel.MSE) for $\bar{q}$, defined as
\begin{equation}
\textrm{Rel.MSE} = \dfrac{\sum_{h=1}^{200} (\bar{q}^{(h)} - Q)^2}{\sum_{h=1}^{200} (\hat{q}^{(h)} - Q)^2},
\end{equation}
where $\bar{q}^{(h)}$ is the value of $\bar{q}$ in simulation $h$, and $\hat{q}^{(h)}$ is the estimate of $Q$ from the sampled data before introduction of missing values
in simulation $h$.

\subsection{Summary of results}

In our initial experiments, MI-GLM crashed when the data
include more than one nominal variable with more than ten categories. This results from
problems estimating the multinomial logistic regressions with many outcome levels. 
Therefore, we consider two types of simulations.  In the first type, we remove the seven variables with more
than ten categories, leaving 28 variables, in order to allow comparisons across all three procedures. In the second type, we include
the seven variables and compare only MI-CART and MI-DPM.
We summarize results using graphical displays and tables. In all plots, MI-GLM is abbreviated as GLM, MI-CART as CART, MI-DPM as
DPM, and the pre-missing data results as NO.  Median coverage rates
across all estimands for each scenario are available in the supplementary material. 
In all tables, the entries are summaries of the Rel.MSEs combined over all relevant probabilities; e.g., 
the 25th, 50th, and 75th percentiles of the Rel.MSEs for all marginal probabilities.
For all results, Monte Carlo standard errors 
for the coverage rates and the reported summaries of the Rel.MSEs are sufficiently small to rule out chance error as explanation for apparent differences in the results.

\subsubsection{Simulations with $n=10000$ and 30\% MCAR}\label{simn=10k30p}

We first create scenarios with $n=10000$ using the MCAR mechanism with the 30\% missingness rate. Including all 28 variables,
we have 98 marginal probabilities, 3343 bivariate probabilities, and 58778  trivariate probabilities. 
Typically, to create $L=10$ completed datasets with a standard desktop computer, MI-GLM runs for about 1 hour and 54 minutes, 
MI-CART runs for about 45 minutes, and MI-DPM runs for about 50 minutes.

\begin{figure}[t]
 \centering
 \includegraphics[width=\textwidth, height = 3.1in, angle=0]{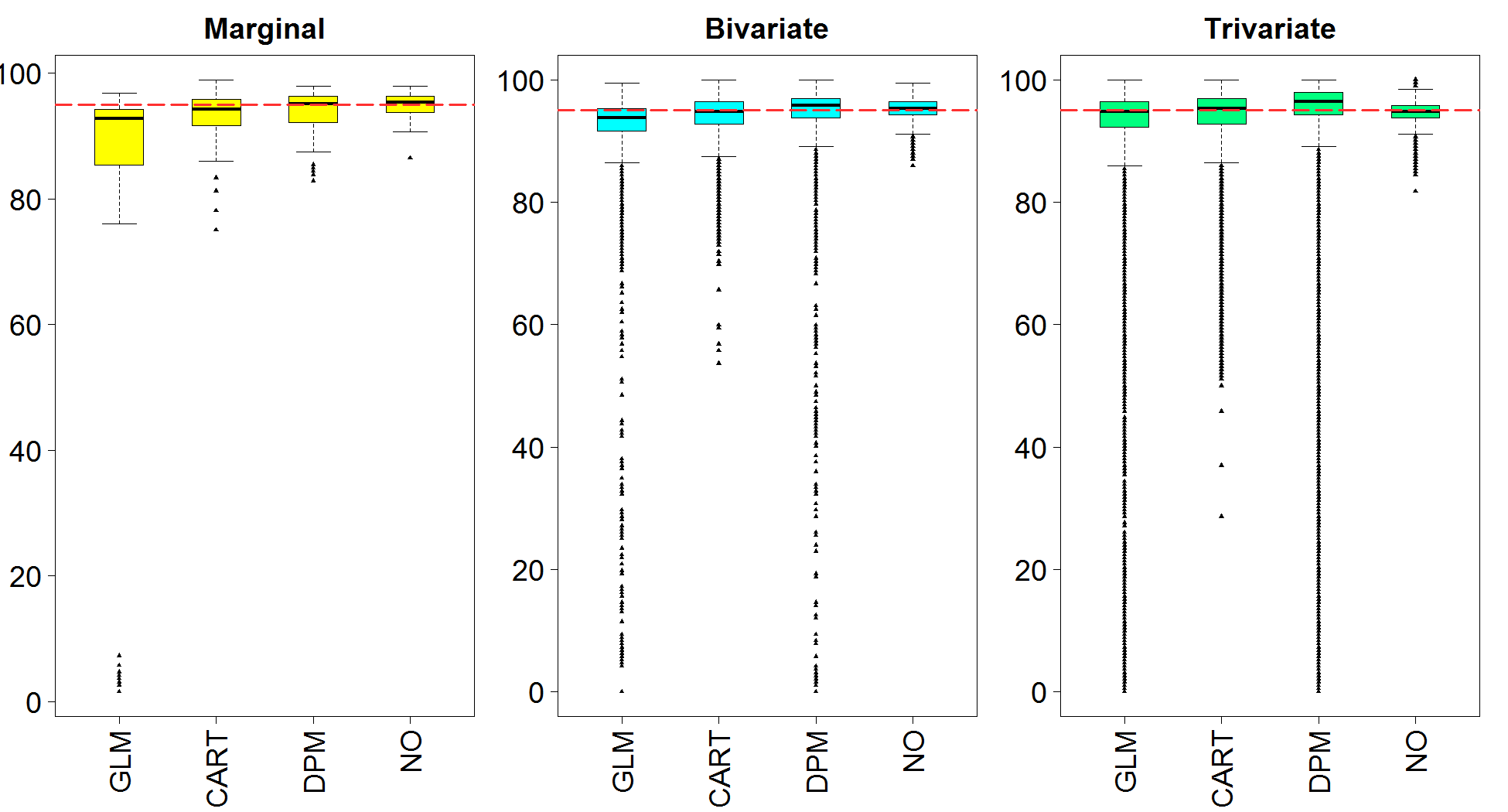}
 \caption{Simulated coverage rates for MI-GLM, MI-CART, MI-DPM, and the pre-missing data intervals when $n=10000$ with 30\% values MCAR. We exclude seven variables with more than ten levels, resulting in $p=28$ variables for imputation and analysis.}
   \label{cov1}
 \end{figure}

Figure \ref{cov1} displays the simulated coverage rates of the 95\%
confidence intervals based on the 28 variables.
For most estimands, all three MI methods result in reasonable coverage
rates.  All three distributions are
skewed to the left, particularly for the bivariate and trivariate
probabilities. Overall among the three MI procedures,  MI-GLM tends to result in the most coverage rates far from the nominal 95\% level.
MI-CART offers the fewest extremely low coverage rates. MI-DPM tends to result in the highest coverage rates, although it also has many low rates for
bivariate and trivariate probabilities.
Across all MI methods, the lowest coverage rates tend to be associated with seven binary variables with marginal probabilities
very close to one in the population. These variables include questions about the presence of bathtubs, refrigerators, running water, sinks, stoves, telephones and flush
 toilets in the households.
 
Table \ref{mse1} displays the values of Rel.MSE for all $\bar{q}$.
The MI-GLM procedure tends to result in the most inaccurate point estimates, and
the MI-CART procedure tends to result in the most accurate point
estimates. Detailed investigations indicate that the differences in performance
reflect differences in biases more than differences in variances.
The largest values in the distributions (e.g., the maxima in Table
\ref{mse1}) usually involve the seven variables with low probabilities
in the population.

\begin{table}[t]
\centering
\caption{Distributions of relative mean squared errors when $n=10000$ and 30\%
  values MCAR. We exclude seven variables with more than ten levels, resulting in $p=28$ variables for analysis between MI-GLM, MI-CART, and MI-DPM.}
   \begin{tabular}{lrrrrrrrrr}
   \hline  \hline
   \multicolumn{1}{c}{}  &   \multicolumn{3}{c}{Marginal} & \multicolumn{3}{c}{Bivariate} & \multicolumn{3}{c}{Trivariate} \\ \cline{2-10}
   & GLM & CART & DPM & GLM & CART & DPM & GLM & CART & DPM \\
     \hline
   Min. & 1.0 & 1.0 & 1.0 & 1.0 & 0.6 & 0.6 & 0.6 & 0.6 & 0.4 \\ 
     1st Qu. & 1.4 & 1.2 & 1.3 & 1.4 & 1.2 & 1.2 & 1.3 & 1.1 & 1.2 \\ 
     Median & 1.6 & 1.4 & 1.5 & 1.6 & 1.4 & 1.4 & 1.5 & 1.3 & 1.4 \\ 
     3rd Qu. & 2.6 & 1.5 & 1.9 & 2.0 & 1.6 & 1.7 & 1.9 & 1.6 & 1.6 \\ 
     Max. & 27670 & 27550 & 6.7 & 39530 & 38530 & 188.4 & 49040 & 47150 & 202.6 \\
      \hline
   \end{tabular}
    \label{mse1}
\end{table}

As a sensitivity analysis, we also remove the seven variables with
probabilities near one, and perform an independent simulation 
of 200 runs on the remaining 21 variables.  We are left with
83 marginal probabilities, 2590 bivariate probabilities
and 37216 trivariate probabilities.  The relevant figure and table are
presented in the online supplementary material.  In short, 
the overall patterns are similar to those in Figure \ref{cov1} and Table \ref{mse1}.
Removing these seven variables also removes most of the extremely low
coverage rates for MI-GLM, making it more competitive with MI-CART
although MI-CART continues to result in slightly better coverage rates overall.  
MI-DPM yields median coverage rates around or slightly above 95\%; however, it continues to
have longer left tails than the other methods for
bivariate and trivariate probabilities.  Removing the seven
variables also removes the extremely large Rel.MSE values seen in
Table \ref{mse1}. The lowest coverage rates and largest Rel.MSEs continue to be associated with probabilities closest to zero or one, 
which is generally the case in all the simulation scenarios we considered.

 \begin{figure*}[t!]
  \centering
     \begin{minipage}[t]{\textwidth}
       \vspace*{\fill}
       \centering
       \includegraphics[width=\textwidth,height=3.1in]{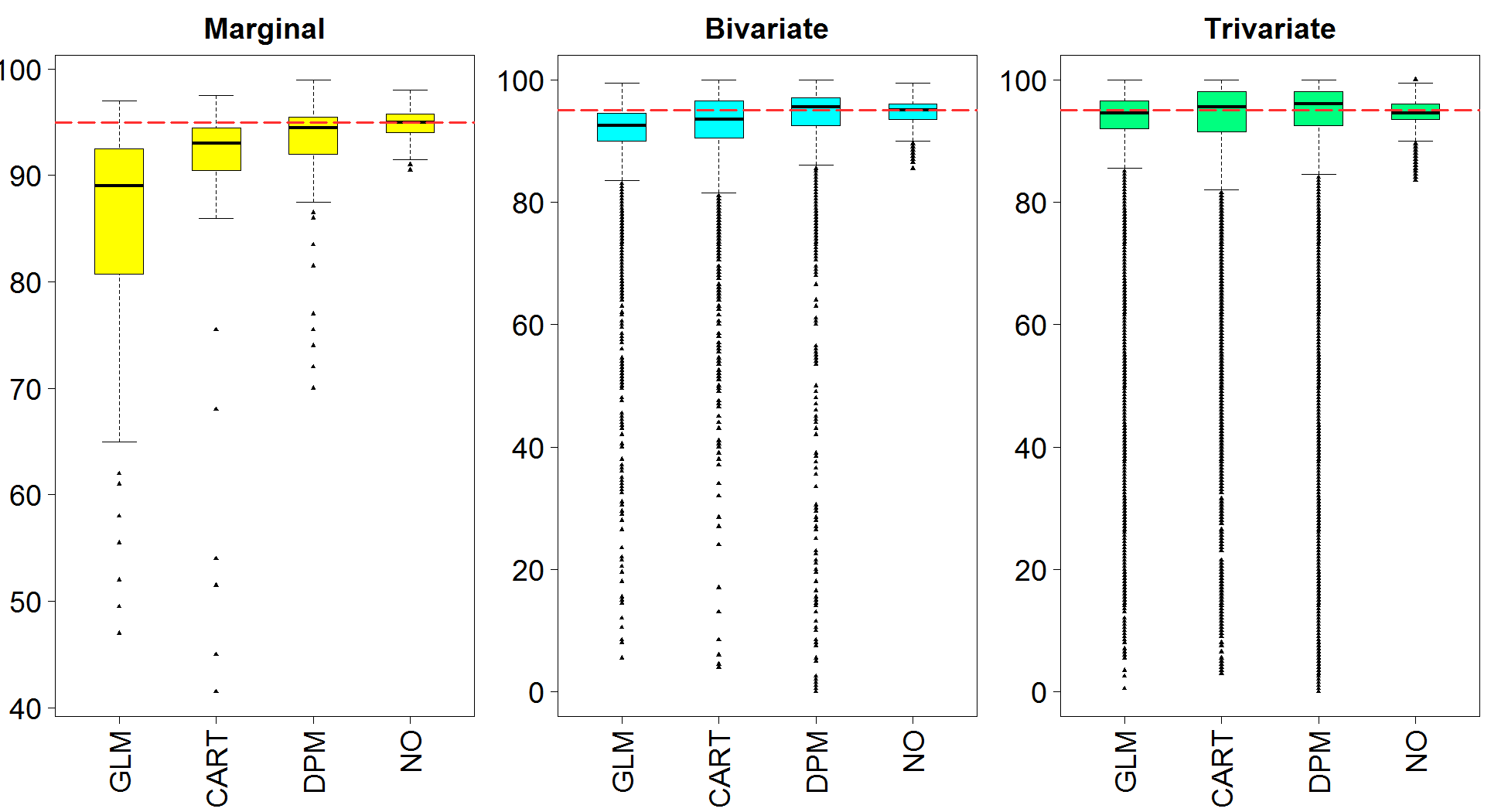}
       \subcaption{Results with 45\% missing data}
       \includegraphics[width=\textwidth,height=3.1in]{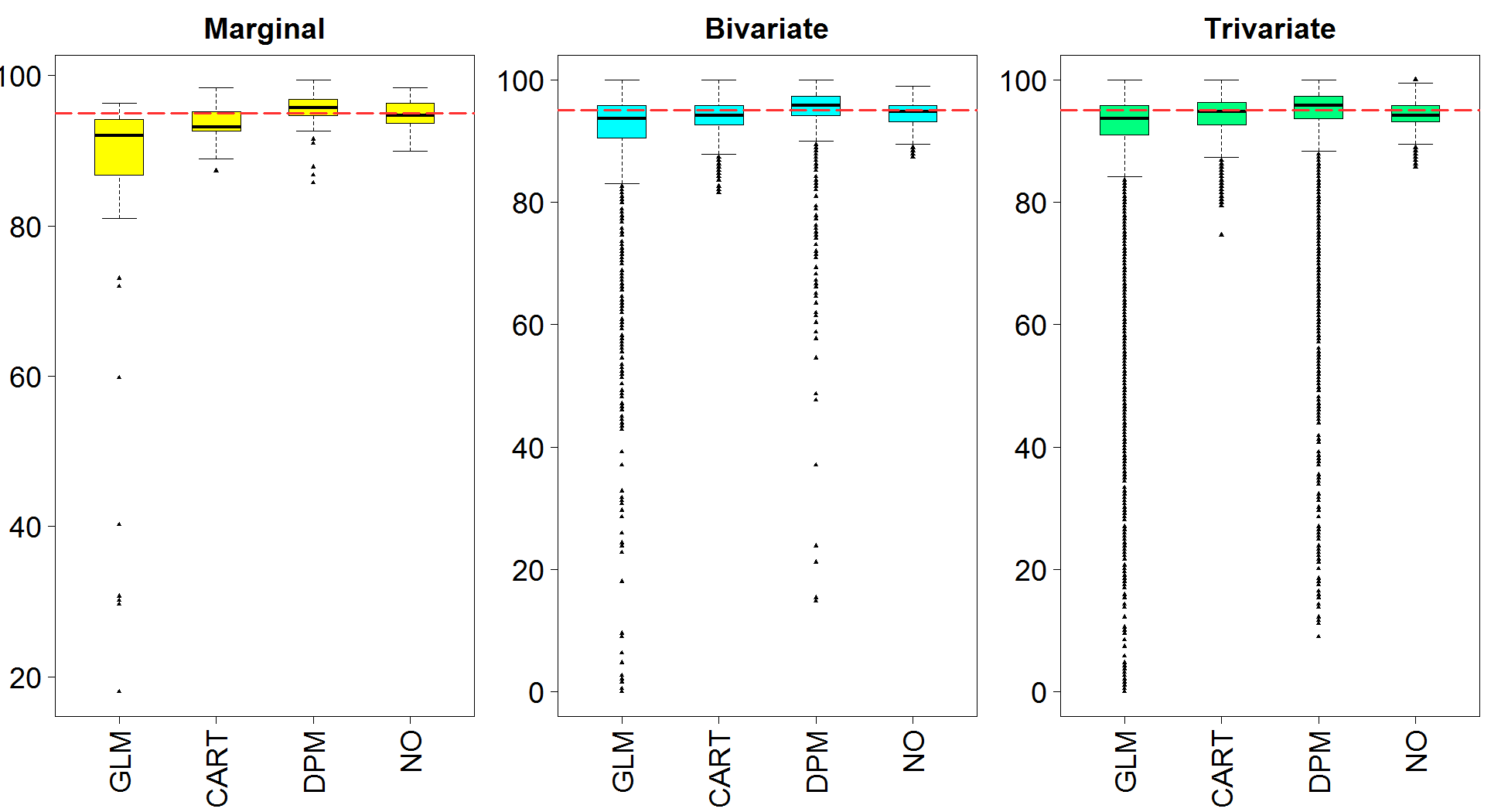}
       \subcaption{Results with $n = 1000$}
     \end{minipage}
     \caption{Simulated coverage rates for  MI-GLM, MI-CART, MI-DPM,
       and the pre-missing data intervals for other MCAR scenarios with $p=21$ variables. Top
         panel is for scenario with $n=10000$ and a 45\% missing data rate, and
         bottom panel is for scenario with $n=1000$ and a 30\% missing data
         rate.}
        \label{cov4}
  \end{figure*}
  
As a final sensitivity analysis, we add in the seven variables with
more than ten categories, and perform an independent set of 200
simulations comparing MI-CART and  MI-DPM only. 
We continue to exclude the variables with probabilities near one. As a
result, the comparison focuses on 28 variables with 177 marginal
probabilities, 
 9049 bivariate probabilities and 180218 trivariate probabilities. To generate $L=10$ completed datasets, 
typically MI-CART takes about 2 hours and 45 minutes, and MI-DPM takes about 55 minutes.
The relevant figure and table are presented in the online supplement.
In short, including the
variables with ten levels does not fundamentally change the
conclusions about MI-CART and MI-DPM.

\subsubsection{Simulations with other MCAR scenarios}\label{45percent}
  
We next consider two other MCAR scenarios to examine whether or not
overall conclusions change when (i) increasing the rate of missingness
and (ii) decreasing the sample size.  For both scenarios, we use
only the $p=21$ variables described in the first sensitivity analysis of
Section \ref{simn=10k30p}; that is, we discard the variables with more
than 10 levels and the variables associated with marginal probabilities very
close to zero or one.

To examine the role of missingness rate, we increase from 30\% to 45\% of values
MCAR independently for each variable.  We continue to use  $n=10000$. 
The results are presented in the top panels of Figure \ref{cov4} and Table \ref{mse4}.
Compared to results with the 30\% missing data rate, we see increased numbers of low coverage
rates for all three MI methods. This results from reduced numbers of observations on which to
estimate the imputation models.  MI-DPM offers the highest
  coverage rates for marginal probabilities.  The distributions of
  coverage rates for the three methods are similar for bivariate  and
  trivariate probabilities, with again MI-DPM tending to yield higher
  coverate rates. As before, MI-CART tends to offer the smallest
  values of Rel.MSE, with MI-DPM a close second.  MI-GLM tends to have
  the largest Rel.MSEs. 

We also consider a simulation with 10\% of values MCAR independently for each variable.
Not surprisingly, with the low missingness rate, for the most part the Rel.MSEs are similar across the three methods, 
although MI-GLM still tends to have the largest Rel.MSEs overall. All three methods have higher coverage rates than in the previous 
scenarios. MI-DPM continues to yield higher coverage rates for most estimands, but also longer left tails, than MI-CART and MI-GLM,  
particularly for bivariate and trivariate probabilities. 
The relevant figure and table are presented in the online supplementary material.

\begin{table}[t]
\centering
\caption{Distributions of relative mean squared errors for  MI-GLM,
  MI-CART, and MI-DPM for other MCAR scenarios with $p=21$ variables. Top
  panel is for scenario with $n=10000$ and a 45\% missing data rate, and
  bottom panel is for scenario with $n=1000$ and a 30\% missing data
  rate.}  
   \begin{tabular}{lrrrrrrrrr}
   \hline  \hline
   \multicolumn{1}{c}{}  &   \multicolumn{3}{c}{Marginal} & \multicolumn{3}{c}{Bivariate} & \multicolumn{3}{c}{Trivariate} \\ \cline{2-10}
   & GLM & CART & DPM & GLM & CART & DPM & GLM & CART & DPM \\
     \hline
 \multicolumn{10}{c}{} \\
&     \multicolumn{9}{c}{Results with 45\% missing data} \\ \
   Min. & 1.2 & 1.1 & 1.0 & 1.0 & 0.7 & 0.5 & 0.6 & 0.5 & 0.4 \\ 
     1st Qu. & 1.8 & 1.4 & 1.4 & 1.8 & 1.4 & 1.5 & 1.6 & 1.3 & 1.4 \\ 
     Median & 2.4 & 1.7 & 1.8 & 2.3 & 1.8 & 1.8 & 2.0 & 1.7 & 1.8 \\ 
     3rd Qu. & 3.2 & 2.0 & 2.1 & 3.1 & 2.3 & 2.4 & 2.6 & 2.3 & 2.5 \\ 
     Max. & 15.1 & 7.1 & 17.2 & 86.8 & 44.8 & 367.0 & 90.5 & 119.8 & 342.8 \\
      
       \multicolumn{10}{c}{} \\
      &     \multicolumn{9}{c}{Results with $n=1000$} \\ 
           Min. & 1.0 & 1.0 & 1.0 & 0.9 & 0.8 & 0.7 & 0.9 & 0.7 & 0.5 \\ 
             1st Qu. & 1.4 & 1.1 & 1.2 & 1.4 & 1.1 & 1.1 & 1.3 & 1.1 & 1.1 \\ 
             Median & 1.9 & 1.3 & 1.3 & 1.6 & 1.3 & 1.3 & 1.5 & 1.2 & 1.2 \\ 
             3rd Qu. & 2.9 & 1.4 & 1.5 & 2.1 & 1.4 & 1.4 & 1.8 & 1.4 & 1.4 \\ 
             Max. & 27.8 & 14.2 & 3.5 & 47.0 & 21.1 & 20.4 & 66.2 & 21.5 & 31.6 \\ 
              \hline
   \end{tabular}
    \label{mse4}
\end{table}

To examine the role of sample size, we return to the 30\% missingness rate and set
$n=1000$.
As evident in the bottom panel of Figure \ref{cov4}, MI-GLM tends to result in the
most coverage rates below the nominal 95\% level and the most
extremely low rates. For most estimands, MI-DPM results in the highest coverage rates
with most of the density concentrated above 95\%, although the distribution has a long left tail as in previous scenarios. For marginal and bivariate probabilities, MI-CART
coverage rates tend to concentrate below 95\% although with a much
shorter left tail than the other methods. As evident in  
Table \ref{mse4}, the MI-GLM procedure still tends to result
in the least accurate point estimates, and the MI-CART and MI-DPM procedures
give reasonably similar performance.
The large values
of Rel.MSE in Table \ref{mse4} again correspond to low probability
events in the population.

\subsubsection{Simulation with MAR scenario}\label{30percentMAR}

Often MAR is a more plausible assumption than MCAR in practice. We therefore design a simulation scenario with MAR, setting 
$n=10000$ and using the 21 variables described previously.  We set
two variables, household type (HHT) which has 5 levels and
tenure of property/home (TEN) which has 4 levels, to be always fully
observed.   Among households with HHT = 1, in each sample we randomly
and independently blank 15\% of values in each of the remaining 19
variables; the corresponding rates for HHT $\in (2, 3, 4, 5)$ are 35\%,
50\%, 50\% and 30\%, respectively. Similarly, for TEN = $(1, 2, 3, 4)$ we
use missing item rates of 40\%, 15\%, 30\% and 5\%, respectively. This
results in around 40\% missing values spread across 
the 19 variables.  We select TEN and HHT as the fully complete
variables because they have 
statistically significant associations with most of the other 19
variables. 

\begin{figure}[t]
\centering
\includegraphics[width=\textwidth, height = 3.2in, angle=0]{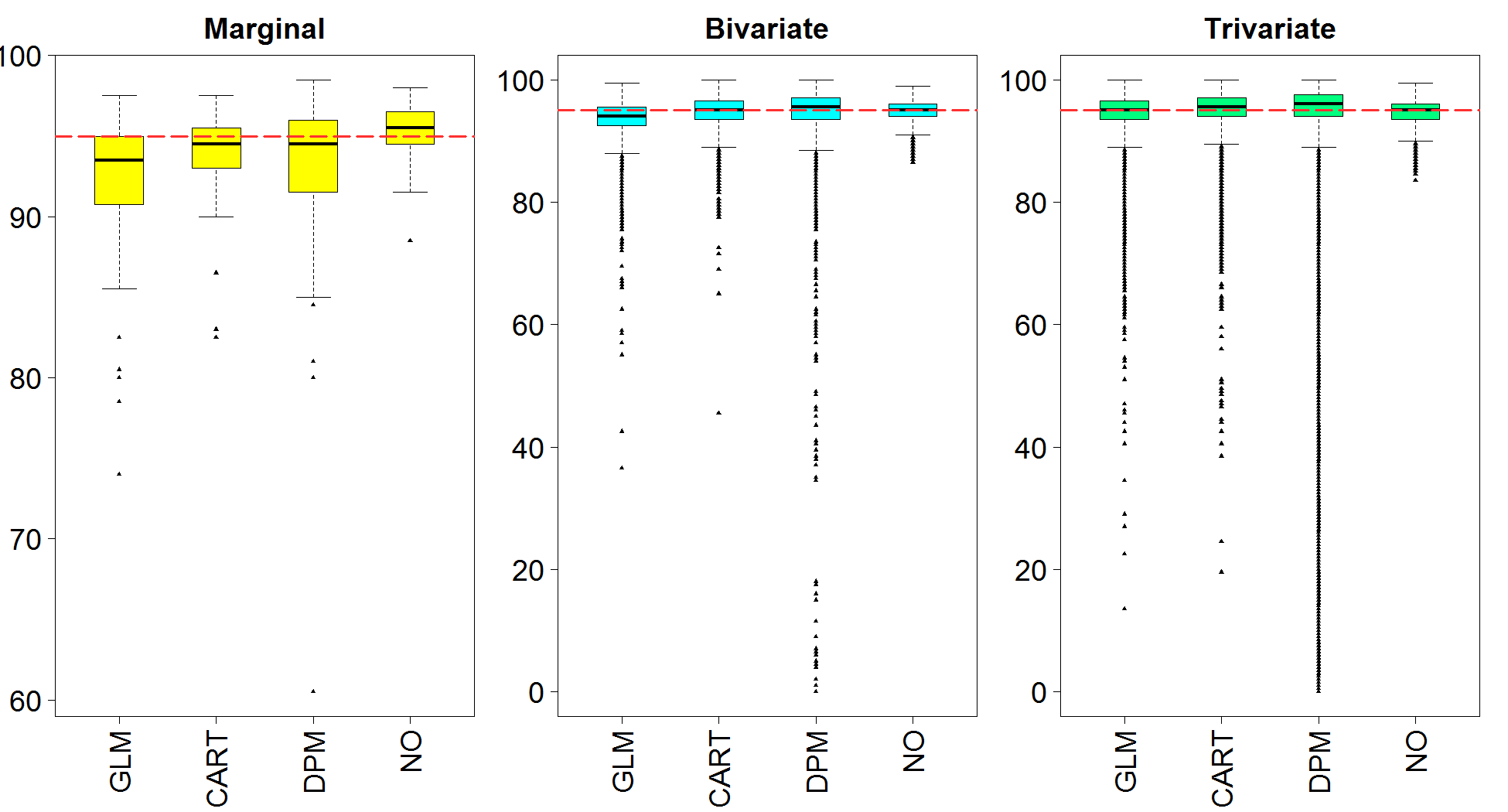}
\caption{Simulated coverage rates for MI-GLM, MI-CART, MI-DPM, and the pre-missing data
  intervals when $n=10000$ and 30\% values MAR. We exclude seven variables with more than ten variables and seven variables with marginal probabilities near one, resulting in $p=21$ variables for imputation and analysis.}
  \label{cov6}
\end{figure}

The results are displayed in Figure \ref{cov6} and Table
\ref{mse6}. The patterns and conclusions from the MAR scenario
are similar to those from the corresponding MCAR scenario.
Overall, MI-CART  tends to result in
the most coverage rates concentrated around 95\% and fewest very low
rates, and the most accurate point estimates.  For most bivariate
and trivariate estimands, MI-DPM still results in coverage rates above
95\%, but it has the longest lower tail, sometimes reaching very low rates. 
MI-GLM results in coverage rates for marginal and bivariate
probabilities that are concentrated slightly below 95\%, but its lower
tail is comparable to that of MI-CART.  
The large values of Rel.MSE in Table \ref{mse6} correspond to low probability
events in the population.

\begin{table}[t]
\centering
\caption{Distributions of relative mean squared errors for  MI-GLM, MI-CART, MI-DPM, and the pre-missing data intervals when $n=10000$ and 30\% values MAR. We exclude seven variables with more than ten variables and seven variables with marginal probabilities near one, resulting in $p=21$ variables for imputation and analysis.}  
         \begin{tabular}{lrrrrrrrrr}
         \hline  \hline
         \multicolumn{1}{c}{}  &   \multicolumn{3}{c}{Marginal} & \multicolumn{3}{c}{Bivariate} & \multicolumn{3}{c}{Trivariate} \\ \cline{2-10}
         & GLM & CART & DPM & GLM & CART & DPM & GLM & CART & DPM \\
           \hline
         Min. & 1.0 & 1.0 & 1.0 & 0.9 & 0.8 & 0.6 & 0.6 & 0.6 & 0.5 \\ 
           1st Qu. & 1.2 & 1.1 & 1.1 & 1.3 & 1.1 & 1.1 & 1.1 & 1.1 & 1.1 \\ 
           Median & 1.5 & 1.3 & 1.4 & 1.5 & 1.3 & 1.4 & 1.4 & 1.2 & 1.3 \\ 
           3rd Qu. & 1.8 & 1.5 & 2.1 & 1.8 & 1.5 & 1.8 & 1.6 & 1.4 & 1.7 \\ 
           Max. & 4.5 & 2.5 & 9.4 & 28.2 & 19.1 & 144.0 & 43.6 & 41.8 & 195.4 \\ 
            \hline
         \end{tabular}
          \label{mse6}
\end{table}

\section{CONCLUDING REMARKS}

The simulation results suggest several general conclusions about the three MI procedures for categorical data.  First,
default applications of MI-GLM with main effects, which are arguably
the most common implementation for multiple imputation, 
appear to be inferior to MI-CART and MI-DPM overall.  These latter
procedures automatically find and model important dependence
structures  
that are missed with default applications of MI-GLM.  Of course, one
could use more complicated predictor functions in the generalized 
linear models, but with high dimensional variables selecting
appropriate sets of interaction effects to include in the conditional models is daunting.
Second, it is difficult to identify a clear winner between MI-CART and
MI-DPM.  Across the simulations, the median coverage rates for MI-DPM
tend to be larger than the median coverage rates for MI-CART, although
both tend to be close to 95\%.  On the other hand, MI-CART tends to
result in fewer very low rates than MI-DPM does, and it tends to result in
smaller relative mean squared errors. Therefore, analysts
  concerned with getting at least nominal coverage rates for most
  estimands, but potentially at the expense of some very low rates, may prefer
  MI-DPM. Analysts willing to accept slightly lower 
  coverage rates for most estimands with potentially lower risk of very low
  rates, in addition to smaller mean squared errors, may prefer
  MI-CART. Third, the results suggest that
one might favor MI-DPM over MI-CART in situations where one needs to lean
on the model more heavily --- the scenarios with smaller sample size and higher missing
data rates in our simulations --- and favor MI-CART otherwise.
Intuitively,
by design the joint model underlying the MI-DPM engine tends to shrink
probability estimates from low count cells towards those for higher
count cells. In modest-sized samples or with high rates of missing data, this
can improve accuracy. The effects of this shrinkage generally decrease
with sample size.  Thus, we conjecture that MI-DPM and MI-CART would
give similar results for much larger $n$; indeed, we found this to be the case in a small number of simulation runs with $n=100,000$.  
Of course, when the sample size is too small, MI-DPM and MI-CART can lack sufficient data to estimate complex relationships accurately, 
highlighting the importance of checking the quality of imputations, e.g., with predictive checks as in \citet{BurgetteReiter2010}, for any method.

As with any simulation study, the conclusions suggested by these results may not generalize to all other settings.  The simulations were based on relatively simple missing data mechanisms, default applications of imputation strategies that were not tuned specifically to the data at hand, and only a subset of the (huge space of) possible estimands. 
For example, it may be that MI-GLM is most effective among the three imputation procedures when  the parameters of interest are a particular GLM that is also used in one of the conditional specifications, although this is not guaranteed to be the case when the specified GLM poorly describes the relationships in the data. 

We considered only data comprising nominal categorical variables.  Of course, many datasets also include ordinal and continuous variables. 
The MI-CART procedure can be easily applied for such mixed data types, but obviously MI-DPM cannot.  One needs a different joint model, such as a general location 
model \citep{Schafer1997a} or the mixture model of \citet{MurrayReiter2016}.  Thus, one should not generalize findings here to mixed data types. 
Additionally, we did not use the ordinal nature of the data when fitting MI-GLM.  Models that do so, like cumulative logit models \citep{Agresti2013}, 
can be more effective than multinomial logistic regressions, provided that their underlying assumptions (e.g., proportional odds) are 
sensible for the data at hand. Proportional odds assumptions do not always match the data distribution, e.g., when the ordinal 
variable has most mass at the lowest and highest values of the variable, so that careful model checking is warranted before using them in 
MI-GLM in lieu of the weaker multinomial logit assumptions.


The simulation results suggest some useful practical steps to improve MI.  First, the performance of the procedures suffered when
variables with probabilities nearly equal to one (or zero) are included in the models. Additionally, occasionally MI-CART, and less frequently
MI-GLM and MI-DPM, generated completed datasets with $b = 0$ for these variables, so that the multiple imputation variance estimator and degrees of freedom 
broke down. Accepting such decreased performance
seems unnecessary, since imputation from marginal distributions are likely to work just as effectively for these variables.
We suggest that analysts remove such variables before imputation. CART initially suffered the most for probabilities close to zero or one before restricting analyses only to quantities with $np > 10$ and $n(1-p) > 10$.
Imposing that restriction and removing those seven variables improved the performance of CART. 
Second, we recommend not considering multinomial regression modeling, as is done in default applications of MI-GLM, when some multinomial variables with missing values have
many levels. We could not get default versions of MI-GLM even to fit in such circumstances.  It may be possible to collapse categories or use fewer predictors in the models, although 
this may sacrifice model quality. This could be problematic when such variables are central to a secondary analysis.
Third, the lack of a clear winner 
suggests benefits for assessing the quality of an imputation method for data akin to those at hand; for example, using repeated sampling simulations with
other, representative data sources or, when large enough, the complete cases. Such evaluations are not
 routinely used in applications, although examples exist
\citep[e.g.,][]{EzzatiEtAl1995, RaghunathanRubin1997, SchaferEtAl1998,
  TangEtAl2005, BuurenEtAl2006}.

\bigskip
\begin{center}
{\large\bf SUPPLEMENTARY MATERIALS}
\end{center}
The supplementary material contains a table describing the variables
used in the simulation, median coverage rates across all scenarios,
and figures for the sensitivity checks in Section \ref{rep_samp_eva}.

\renewcommand{\refname}{REFERENCES}
\printbibliography
\end{document}